\documentclass[doublecol]{epl2} 

\newcommand{\be}{\begin{equation}}
\newcommand{\ee}{\end{equation}}
\newcommand{\bea}{\begin{eqnarray}}
\newcommand{\eea}{\end{eqnarray}}

\newcommand{\half}{\mbox{\small $\frac{1}{2}$}}

\newcommand{\eexp}{\mbox{e}^}
\newcommand{\bra}{\left\langle}
\newcommand{\ket}{\right\rangle}

\newcommand{\eqref}[1]{(\ref{#1})}
\newcommand{\beq}{\begin{eqnarray}}
\newcommand{\eeq}{\end{eqnarray}}
\newcommand{\hide}[1]{}
\newcommand{\ba}{{\bar \alpha}}

\title{Zero temperature geometric 
spin dephasing on a ring in presence of an Ohmic environment}
\shorttitle{Geometric spin dephasing on a ring}

\author{B. Horovitz\inst{1} \and P.  Le Doussal\inst{2}
\and  G. Zarand\inst{3,4}}

\institute{                    
  \inst{1} Department of Physics, Ben Gurion university, Beer Sheva
84105 Israel  \\
  \inst{2} CNRS-Laboratoire de Physique Th{\'e}orique de
l'Ecole Normale Sup{\'e}rieure, 24 rue Lhomond,75231 Cedex 05,
Paris France 
\\
   \inst{3}
Freie Universi\"at Berlin, Fachbereich Physik, Arnimallee 14, D-14195 Berlin, Germany
\\
\inst{4}
Theoretical Physics Department, 
Budapest University of Technology and Economics, Budafoki ut 8, 
H-1521, Hungary
}


\abstract{
We study zero temperature spin dynamics of a particle confined to a ring in presence of spin orbit coupling and Ohmic electromagnetic fluctuations.
We show that the dynamics of the angular position $\theta(t)$ are decoupled
from the spin dynamics and that the latter is mapped to certain
correlations of a spinless particle. We find that the spin
correlations in the $z$ direction (perpendicular to the ring) are
finite at long times, i.e. do not dephase. The parallel (in plane) 
components for
spin $\half$ do not dephase at weak dissipation but they probably
decay as a power law with time  at strong dissipation.
}

\begin{document}

\maketitle

\section{Introduction}

Due to recent advances in semiconductor technology, it became
 possible to  isolate and manipulate spins of individual
electrons \cite{Elzerman,Nowack}. For efficient spin manipulation, however, slow
spin decay is needed. Spin decay in mesoscpopic devices
is generated by two major sources: Hyperfine interaction with
nuclear spins \cite{LossGlazman} is responsible for spin decay in most
materials. However, spin-orbit (SO) coupling can
also induce spin relaxation, and
under certain conditions,
phonon-\cite{Nazarov}
or electromagnetic field-induced SO relaxation~\cite{SanJose}
can dominate the decay\cite{meunier,Amasha}.
As shown in Ref.~\cite{SanJose}  two-photon (or two-phonon) processes
lead to {\em geometrical spin relaxation} even in the absence of external
field and, as pointed out recently, this mechanism
can become even dominant in hole-doped systems~\cite{Gerardot,Pascal}.

Here  we make an attempt to understand, whether the above-mentioned
geometrical spin relaxation can survive even at $T=0$ temperature.
Although Ohmic electromagnetic fluctuations
were found to lead to a vanishing spin
relaxation rate at $T=0$~\cite{SanJose},
the results of Ref.~\cite{SanJose} are not
conclusive, since they allow for {\em non-exponential} relaxation, common
in Ohmic systems. To address this issue more rigorously, we consider a
ring geometry. Studying a ring is, however, not of pure theoretical interest;
high quality semiconductor
rings \cite{fuhrer,ihn} can in fact also be used as quantum spin qubits
\cite{ihn}, and usefulness of these devices
depends on spin dephasing, a topic under active experimental study
\cite{meunier,Amasha,Gerardot,pfund,koppens}.

There are two types of spin-orbit coupling in
two-dimensional electron systems: the
Rashba interaction induced, e.g.,  by an electric field
perpendicular to a two-dimensional  (2D) layer~ \cite{rashba},
and the Dresselhaus coupling induced by bulk inversion
asymmetry~\cite{dressel}. Our aim
is to study, how these couplings influence spin coherence for an
electron confined to a ring, in the presence of  Ohmic fluctuations.
We shall first derive the appropriate Hamiltonian for a confined electron, and
show that the presence of the spin does not influence the orbital motion of
the confined electron, which is  governed exclusively by fluctuations of the
external electric field. The dynamics of the spin, on the other hand, is
determined by the orbital motion of the electron, and  has a
topological character. We find that for weak dissipation the spin does
not dephase, but certain spin components are reduced
by fluctuations.  For strong dissipation, however, we find that certain
components of the spin probably relax even at $T=0$ temperature, 
due to the disordering
of the orbital degrees of freedom~\cite{Spohn}. The relaxation
we find is, however, not exponential but of a power law,
typical of confined particles at 
temperature $T=0$~\cite{schoen,etzioni}.

\section{Hamiltonian}

Let us start by projecting a 2D spin-orbit Hamiltonian on a ring, a procedure
which is not entirely trivial \cite{meijer,sheng}. In addition to the
kinetic terms, the 2D Hamiltonian
consists of a potential $V_0(r)$ that confines the particle
to a ring of radius $R\pm \delta R$, with $\delta R \ll R$.
We write the total Hamiltonian in polar coordinates as
${\cal H}_0+{\cal H}'$ where
\beq
{\cal
H}_0&=&-\frac{\hbar^2}{2m_e}\left[\frac{\partial^2}{\partial
r^2}+\frac{1}{r}\frac{\partial}{\partial r}\right]
+V_0(r)\;,\\
{\cal
H}'&=&\frac{p_{\theta}^2}{2m_er^2}+\alpha_0({\bf S}_xp_y-{\bf S}_yp_x)+
\beta_0({\bf S}_xp_x-{\bf S}_yp_y)
\;.
\nonumber
\eeq
Here $p_{\theta}=-i\hbar\partial/\partial_{\theta}$, ${\bf S}$ are
spin operators, $m_e$ is the electron mass and $p_x$ and $p_y$ denote the $x$ and $y$ components of the
momentum.  The $\alpha_0$ term is the Rashba coupling while
$\beta_0$  denotes the Dresselhaus coupling.
Labeling the radial eigenstates of ${\cal H}_0$ by $|n\rangle$
and their energies by $E_n$, our aim is to project ${\cal H}'$
on the subspace, $|0\rangle$, while keeping terms up to order $O(\delta R)$,
a procedure that involves some subtleties.
First we rewrite ${\cal H}'={\cal H}_1+{\cal H}_2$ by introducing
${\bf S}_r^{\pm}\equiv  \cos\theta \; {\bf S}_x \pm \sin\theta\; {\bf S}_y$
and
${\bf S}_{\theta}^{\pm}\equiv \cos\theta \;{\bf S}_y \mp \sin\theta\;{\bf S}_x$,
\beq
{\cal H}_1&=&\frac{p^2_{\theta}}{2mr^2}+
\frac{\alpha_0}{2r}\{{\bf S}_r^+,p_{\theta}\}-\frac{\beta_0}{2r}\{{\bf S}_{\theta}^-,p_{\theta}\}
\;,
\nonumber\\
{\cal H}_2&=&i\alpha_0\hbar{\bf S}_{\theta}^+
(\partial_r+\frac{1}{2r})-i\beta_0\hbar{\bf S}_r^-
(\partial_r+\frac{1}{2r})
\;.
\eeq
As noticed by  Meijer et al. \cite{meijer},
for any state $\psi(r)$ that is radially localized near ${R}$, one has
$\langle\psi |
2\partial_r+\frac{1}{r}|\psi\rangle=0$.
Therefore, to first order in the SO coupling, ${\cal H}_2$ does not give a
contribution to the projected Hamiltonian.
Nevertheless, as previously  overlooked, ${\cal H}_2$ cannot be ignored:
localization on a scale $\delta R$ implies $\partial_r\sim
1/\delta R$ and hence 2nd order perturbations in ${\cal H}_2$
do give a contribution, $\sim {\cal H}_2^2/E_n=O(1)$, since $E_n\sim 1/(\delta
R)^2$.  The next order contributions scale as ${\cal H}_2^3/E_n^2=O(\delta
R)$, and vanish in the limit $\delta R\to 0$, similar to all higher order
terms in the perturbation series.

Perturbation theory to 2nd order
yields therefore the projected spin and angle dependent effective Hamiltonian
\beq
{\cal H}_{\rm ring}=\bra0| {\cal H}_1|0\ket -\sum_{n\neq 0}\frac{
\bra 0|{\cal H}_2|n\ket
\bra n|{\cal H}_2|0\ket
}{E_n-E_0} + {\cal O}(\delta R)\;.
\label{eq:PT}
\eeq
The sum in Eq.~(\ref{eq:PT})
can be evaluated analytically by making use of a sum
rule~\cite{kohn,future}, and
%
%
%
%
the second term of Eq.~(\ref{eq:PT})
simply becomes $ \half m_e(\alpha_0{\bf S}_{\theta}-\beta_0{\bf S}'_r)^2$.

Introducing  the vector ${\bm h}(\theta)$ via
${\bm h}\equiv
(\alpha\cos\theta-\beta\sin\theta,\alpha\sin\theta-\beta\cos\theta)$,
and with the dimensionless Rashba and Dresselhaus couplings defined as
 $\alpha\equiv m{R}\alpha_0$ and $\beta\equiv m{R}\beta_0$,
we can finally rewrite our effective ring Hamiltonian in the
$\delta R\to0$ limit as
\beq
{\cal H}_{ring}=\frac{\hbar^2}{2m_eR^2}[p_{\theta}+{\bm
h}(\theta)\cdot{\bf S}]^2 \;. \label{model1}
\eeq
We remark, in particular, that the term $\sim \alpha\beta\sin 2\theta$ in the
effective Hamiltonian of Ref.~\cite{sheng} is exactly canceled
by the 2nd order terms. As a consequence, Eq.~\eqref{model1}
possesses a conserved "momentum",
$\hat Q \equiv p_{\theta}+{\bm h}(\theta)\cdot{\bf S}$.
%

\section{Spectrum}

The eigenstates and eigenenergies of \eqref{model1}
can be analytically computed 
for $\beta=0$, when the system is rotationally
invariant and therefore $J_z=p_{\theta}+S_z$ is also conserved.
The Hamiltonian can then be written as
\beq
{\cal H}_{\rm ring}=&\frac{\hbar^2}{2m_eR^2}[J_z-{\bm
n}(\theta)\cdot{\bf S}\sqrt{1+\alpha^2}\,]^2\;,
\label{H_ring2}
\eeq
with ${\bm n}(\theta)=(-h_x(\theta),-h_y(\theta),1)/\sqrt{1+\alpha^2}$
a unit vector.
The energy spectrum and the eigenvalues can then easily be found
by constructing
common eigenstates of the two commuting operators, $J_z$ and ${\bm
  n}(\theta)\cdot{\bf S}$.
For $S=1/2$,   ${\bm n}(\theta)\cdot{\bf S}$ and
 $J_z$ have eigenvalues ${\bm n}(\theta)\cdot{\bf S}=\sigma/2$
and $J_z=m+\sigma/2$, respectively, with $\sigma=\pm$
and $m$ an integer. The spectrum is $\epsilon_{m,\sigma} = \frac{1}{2 m_e R^2}
\bigl[m+\sigma ( \half-\frac{1}{2} \sqrt{1+\alpha^2})\bigr]^2$, and
the eigenstates are of the form
$\int_{\theta} \frac{e^{i m \theta}}{\sqrt{2\pi}} |\theta\rangle \otimes | \pm {\bm
  n}(\theta)\rangle$, with $| \pm {\bm n}\rangle$ denoting spin coherent
states,  defined through the usual relation,  ${\bm \Omega} \cdot{\bf S}  |{\bm \Omega}\rangle = S|{\bm
  \Omega}\rangle$ \cite{Assa}. The wave functions can be explicitly expressed as
\begin{eqnarray}
\psi_{m,+}(\theta)&=&\eexp{im\theta}
\left(   \begin{array}{cc}
      \cos \frac{\ba}{2},
       -\eexp{i\theta}\sin
      \frac{\ba}{2}  \end{array} \right)\;,
\nonumber\\
\psi_{m,-}(\theta)&=&\eexp{im\theta}
\left(   \begin{array}{cc}
      \eexp{-i\theta}\sin \frac{\ba}{2},
      \cos \frac{\ba}{2}  \end{array} \right)\;,
\label{eq:spinor}
\end{eqnarray}
with $\ba$ defined as $\ba\equiv \arctan (\alpha)$.
The states $\psi_{\pm m,\pm}$ are related by time reversal,
and their energies equal,  $E_{m,+}=E_{-m,-}$.
For $\alpha<\sqrt{3}$  the
ground state has $m=0$.\footnote{
Here we used the phase convention of
Ref.~\cite{Assa}. This construction can
be generalized for larger spins.}

\section{Dissipation}

Having understood the properties of an isolated ring,
we now couple  the motion of the particle
to the coordinate $\xi$ of a dissipative environment,
i.e., we consider the total Hamiltonian
as ${\cal H}={\cal H}_{ring}+V(\theta,\xi)$.
Throughout most of this paper we shall assume that
$V(\theta,\xi)$ describes the coupling to
a Caldeira-Leggett (CL)
environment, appropriate for small rings in an Ohmic (metallic)
environment.\footnote{In a dirty metal environment, e.g.,
 one needs the ring's radius to be smaller than the mean free path.}
Then $\xi$ represents the random force generated by the environment,
 $V=\xi_-\eexp{i\theta}+\xi_+\eexp{-i\theta}$, and
the $T=0$ Fourier transform of the environment correlations is
$\langle \xi_-\xi_++\xi_+\xi_-\rangle_\omega =\hbar^2\;\eta|\omega|$, with
$\eta$  the dimensionless friction coefficient.

\begin{figure}[t]
\centering
\includegraphics[width=2.5in]{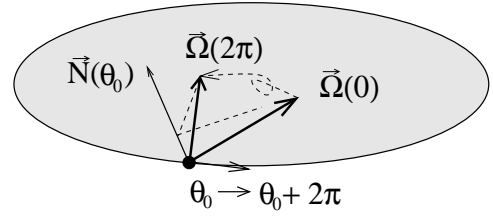}
\caption{Evolution of the  spin (coherent state) while the electron
makes a circle, $\theta_0\to\theta_0+2\pi$. The initial
state is rotated around an axis ${\bf N}(\theta_0)$ by an angle
  $\Gamma_{\theta_0}$.} \label{fig:ring}
\end{figure}

The corresponding equations of motion for $\theta(t)$ are
\beq
\dot{\theta}=\frac{p_{\theta}+{\bm h}\cdot {\bf S}}{m_eR^2}\;,\qquad
\ddot{\theta}=-\frac{1}{m_eR^2}\partial_{\theta}V(\theta,\xi)\;.  \label{spindec}
\eeq
Hence, as a consequence of the simple form of ${\cal H}_{\rm ring}$, 
Eq.~\eqref{model1},
 the dynamics of $\theta$ in the dissipative environment
are not affected by the spin-orbit couplings.
This decoupling allows us to describe the $\theta(t)$ evolution by a
path integral, where for  each trajectory the spin dynamics
follow from Eq.~(\ref{model1})
\beq
\frac{d{\bf S}}{dt}=\dot{\theta}\;{\bm h}(\theta)\times {\bf S} \qquad\Rightarrow
\qquad \frac{d{\bf S}}{d\theta}={\bm h}(\theta)\times {\bf S}\;.
\label{eq:spin_equation}
\eeq
Viewing $\theta$ as a "time" variable, these dynamics correspond to a spin
precession  around a "time" dependent magnetic field ${\bf h}(\theta)$.
Note that switching to  "Schr\"odinger" picture, 
the spin coherent states have a simple $\theta$ evolution, too.
Apart from a phase, they evolve as 
$|{\bf \Omega}(\theta)\rangle$,  where ${\bf \Omega}(\theta)$ 
is the vector solution
of (\ref{eq:spin_equation}), i.e. 
$\frac{d{\bf \Omega}}{d\theta}={\bm h}(\theta)\times {\bf \Omega}$.
In particular, the vector ${\bf n}(\theta)$ can also be shown to satisfy 
this equation.

In terms of the spin operators, Eq.~\eqref{eq:spin_equation} is solved
as a simple linear mapping $S_i(\theta) = R_{ij}(\theta,\theta_0) S_j$,
with  $R_{ij}(\theta,\theta_0)$ a rotation matrix.
The rotation matrix $R_{2 \pi}(\theta_0)=R(\theta_0+2 \pi,\theta_0)$,
corresponding to the particle going once around the ring by $2 \pi$, is of
special interest. We denote by the unit vector ${\bf N}(\theta_0)$ its 
axis of rotation and by $\Gamma_{\theta_0}$  the corresponding 
rotation angle (see Fig.~\ref{fig:ring}). In particular, for 
 $\beta=0$ we find that the angle 
$\Gamma_{\theta_0}$ is independent of the initial value, 
$\Gamma = 2\pi (1-\sqrt{1 + \alpha^2})$, and is typically 
incommensurate with $2\pi$.


%


%

\section{Mapping to a spinless system}

For a given evolution, $\theta_0\to \theta$,
we can obtain the evolution of the spin part of the wave function
from Eq.~\eqref{eq:spin_equation}, which is described by
a unitary operator, $U_{\rm spin}(\theta,\theta_0)$.
Here the Hamiltonian to describe the $\theta$ ("time") evolution
of the spin is 
$H_s=
\mathbf{h}(\theta) \cdot \mathbf{S}$.
We proceed to study the case $\beta=0$. Then, as 
in the standard NMR rotating field problem,
 the spinor transformation $\psi'\equiv 
e^{ i (\theta-\theta_0) S_z} \psi$ to the 
"rotating frame" cancels the ''time'' ($\theta$) dependence,
and amounts in replacing  
$H_s\to \mathbf{h}(\theta_0) \cdot \mathbf{S} -S_z = - \sqrt{1+\alpha^2}\; 
\mathbf{n}(\theta_0) \cdot \mathbf{S}$. 
For $S=1/2$ this leads to the evolution operator, 
$ U_{\rm spin}(\theta,\theta_0) = e^{ - i \frac{\theta-\theta_0}{2} \sigma_z}
e^{ i \sqrt{1+\alpha^2} \frac{\theta-\theta_0}{2} \mathbf{n}(\theta_0)
  \cdot \mathbf{\sigma}}$.
Using now the expression of $\hat Q$ we find that 
the $\theta$ evolution of the spin states has a particularly
simple form
\beq
U_{\rm spin}(\theta,\theta_0)\psi_{m,\pm}(\theta_0)
=\eexp{iq_{m,\pm}(\theta_0-\theta)}\;\psi_{m,\pm}(\theta)\;,
\label{simple}
\eeq
where $q_{m\;\pm}=m  \pm\half\mp\half \sqrt{1+\alpha^2}$ denote the
eigenvalues of the momentum $\hat Q$.
After a $2\pi$ rotation the state $\psi_{m,\pm}$
picks up an  incommensurate phase,  $2\pi q_{m,\pm}$.
Note that the semiclassical evolution involves a similar 
incommensurate angle, $\Gamma$, as
discussed below Eq.~\eqref{eq:spin_equation}  (see also Fig. 1).

Making use of the decoupling of orbital and spin degrees of freedom, 
we can construct a mixed path integral formalism 
(to be detailed in Ref.~\cite{future}), where the spin is
treated in an evolution operator formalism,
while the orbital motion of the particle
is developed in a path integral formalism. 
The full  evolution for a given
environment history is then obtained as:
\beq
&&\psi_{m,\pm}(\theta_t,t)=\sum_n\int_0^{2\pi}d\theta_0\int_{\theta_0}^{\theta_t+2\pi n}
{\cal D}\theta \;
\nonumber
\\
&&\phantom{nnnn}\eexp{i S_P(\theta,\xi)}
\; U_{spin}(\theta_t+2\pi n,\theta_0)\psi_{m,\pm}(\theta_0)\;.
\label{spin_propag}
\eeq
Note that $\theta$ in this equation is a non-compact variable,
and an additional integration over the environment configurations
has to be carried out in the end. Importantly, 
the action $S_P(\theta,\xi)= \int_0^t [ \half m_er^2{\dot \theta}^2 - V(\theta,\xi)]$
describes a particle on the ring in the presence of dissipation for a
given environment history, and is independent of the spin evolution.

For $\beta=0$ we can make use of Eq.~\eqref{simple} and 
obtain a particularly simple path integral representation for 
the spin evolution.
Consider  spin correlations with an initial density matrix
$|\sigma\rangle\langle \sigma|$ built from  one of 
the two Kramers degenerate ground
states of $m=0$ and $\sigma=\pm$, having momenta
$\hat Q = q_{0,\pm}=\pm G$ with $G=\half-\half\sqrt{1+\alpha^2}$. Using Eqs.~\eqref{spin_propag} and \eqref{simple}
for the  forward and backward spin evolutions,
we find an exact mapping of
the spin correlations onto a superposition of 
equilibrium correlations of spinless particles on a
ring with a flux $\Phi=\pm G$ (in units of quantum flux):
\beq
P_{a,\Phi}(t_{21})=\langle \eexp{-ia\theta(t_2)}\eexp{ia\theta(t_1)}\rangle_{\Phi}\;.
\label{eq:P_a,q}
\eeq
Here again, $\theta(t)$ is a non-compact variable within $(-\infty,\infty)$
to be used within the path integral representation of the spinless
problem. Note that the bath still couples to $\eexp{\pm i\theta}$
hence we expect that \eqref{eq:P_a,q} depends only on the noninteger part of
$\Phi$.  For  $\langle S_x(t)S_x(0)\rangle$ we obtain the following
identity 
for an initial density matrix, $|+\rangle\langle +|$,
\beq
C_{++}^x(t)&=&\frac{1}{4}\sin^2\ba(P_{1,G}(t)+P_{-1,G}(t))
\label{C^x_++}
\\
&+&\cos^4\frac{\ba}{2}P_{-2G,G}(t) 
+\sin^4\frac{\ba}{2} P_{2-2G,G}(t)\;.
\nonumber
\eeq
For $C_{--}(t)$
the same result holds with all subscripts of
$P_{a,Q}$ reversing sign. For the $S_z$  correlations, on the other hand,
 we obtain
\beq
C_{++}^z(t)=\cos^2\ba + P_{-1-2G,G}(t)\sin^2\ba\;,
\label{C^z_++}
\eeq
and for $C_{--}^z(t)$ the same holds with $P_{1+2G,-G}$. 
Notice that the degeneracy point,
$\alpha=\sqrt{3}$, corresponding  to fluxes $\Phi=\pm\half$ 
represents a special case, and is not  studied here.

While the correlation function of the $z$-component of the spin,
$C^z(t)$, obviously contains a constant
non-decaying piece, the correlation function $C^x(t)$ contains
only phase correlation functions $P_{a,\Phi}$ with $a\ne 0$. There is some evidence that 
these correlations decay in time. In particular $P_{1,0}\sim 1/t^2$ from the XY lattice model
\cite{Spohn} and from small $\eta$ expansion \cite{hl}. Correlations
with incommensurate $a$ were studied in a related system of
dissipative  Josephson junctions \cite{lukyanov}, and 
found to decay algebraically.
Further evidence is for large $\eta$, as discussed below. To further appreciate these correlations
we have
evaluated the path
integrals in \eqref{eq:P_a,q} analytically for $\eta=0$, and
surprisingly, we find $P^{\eta=0}_{\mp 2G,\pm G}(t)=1$.
As a consequence, for $\eta=0$ the
correlation function $C^x$  contains a piece which does not
oscillate.
As discussed below, though reduced, this part seems to
survive for very weak dissipation, $\eta\ll 1$, while it apparently
decays algebraically for strong dissipation, $\eta\gg 1$.

\section{Strong dissipation limit}

In the strong dissipation limit, $\eta\gg 1$, we can describe
the evolution of the phase through a Langevin
equation, and an expansion
in $1/\eta$ is possible~\cite{schoen,etzioni}.
In this limit, all correlation functions $P_{a,q}$ with $a\ne 0$
are found  to decay algebraically. Large $\eta$ perturbation theory
yields that $P_{a,\Phi} \sim t^{-a^2/\pi \eta}$ and the $x$-component of
the spin also decays algebraically, while the $z$-component
remains finite and does not decay. This result,  however, holds only for
up to times $\ln t < O(\eta)$ beyond which effects of renormalization
of $\eta$ cannot be neglected. In a recent work \cite{etzioni} we have shown that in presence of
a weak DC electric field there is a critical $\eta_c=1/2\pi$ such that $\eta>\eta_c$ flows to $\eta_c$ which would indicate $P_{a,\Phi} \sim t^{-2 a^2}$. In some sense the fluctuating spin corresponds to a time dependent flux, i.e. an electric field, though the correspondence is not precise.


\section{Weak dissipation}

The rather different behavior of $S_{x,y}$ and $S_z$ should
already be manifest in the  weak dissipation limit, where
we can perform perturbation
theory  in the strength of the dissipation, ${\eta}$.
To do perturbation theory, we  restrict ourselves to the
case $S=1/2$ and $\beta=0$, and use Abrikosov's pseudofermion method
to represent each spinor $\psi_{m\sigma}$ of \eqref{eq:spinor} by a
pseudofermion operator,  $f_{m\sigma}$.
In this language the ring Hamiltonian becomes
${\cal H}_{\rm ring}=\sum_{m,\sigma} \epsilon_{m\;\sigma}\;
f_{m\;\sigma}^{\dagger}f_{m\;\sigma}$, while the interaction
is expressed as
\begin{eqnarray}\label{hpf}
V = \sum_{m,\sigma} \left( \xi_-f_{m\;\sigma}^{\dagger}f_{m-1\;\sigma}+
\xi_+ f_{m\;\sigma}^{\dagger}f_{m+1\;\sigma}
\right)\;,
\end{eqnarray}
and standard field-theoretical methods can be used to evaluate
physical quantities.
A renormalization group analysis of
 the vertex function and the pseudofermions'
self-energy reveals that, although ultraviolet logarithmic divergencies appear in
both quantities, they cancel and the dissipation parameter $\eta$ 
is in leading order, nevertheless,
{\em exactly marginal}, and the mass of the particle remains also
unrenormalized \cite{future}. 

\begin{figure}[t]
\centering
\includegraphics[width=2in]{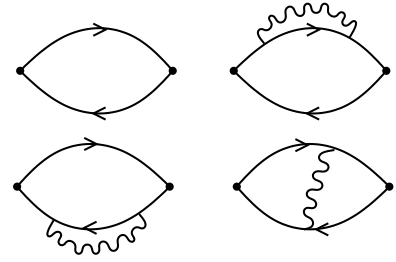}
\caption{
Zero and leading order ($\sim\eta$) corrections to the spin susceptibility.
Continuous lines and wavy lines represent the pseudofermion and
bosonic propagators, respectively,  while the dots represent spin vertices.
Logarithmic divergencies in the three diagrams above exactly cancel.
} \label{diagrams}
\end{figure}

In this perturbative regime,
fingerprints of a non-exponential spin decay should appear in the
susceptibility, $\chi$, which, in the absence of spin decay, should contain
a  Curie part. To compute $\chi$, we first
express the impurity spin operator in terms of pseudofermions as
\beq
\mathbf{S}_{i} = \sum_{m,\sigma,m\prime,\sigma\prime}
{\cal
  S}^i_{m,\sigma,m\prime,\sigma\prime}f_{m\;\sigma}^{\dagger}f_{m\prime\;\sigma\prime}\;,
\label{pseudofermion_spin}
\eeq
with the matrix elements simply determined from the wave
functions \eqref{eq:spinor}, as ${\cal S}^i_{m,\sigma,m\prime,\sigma\prime}
= \langle \Psi_{m,\sigma}|\mathbf{S}_{i} | \Psi_{m,\sigma}\rangle $.
The leading corrections to $\chi$ are shown
in Fig.~\ref{diagrams}. Although all corrections shown contain logarithmic
ultraviolet singularities, remarkably, all these singularities
exactly cancel, and  one  finally
obtains  just a finite renormalization of the perpendicular Curie
susceptibility 
\beq
\chi_{x,y}= \frac{\cos^4({\ba}/2)}{4 T}\Bigl[1-\frac{\eta}{2\pi\;} [\frac{1}{G}
\ln \Bigl( \frac{1 + 2 G}{1-2G} \Bigr) - 4] + O(\eta^2)
\Bigr]\;.
\nonumber
\eeq
The prefactor $\cos^4({\ba}/2)$ originates from the coefficient of
$P_{-2G,G}(t)$ in Eq.~\eqref{C^x_++}, and accounts for
$g$-factor renormalization
in the isolated ring. The correction $\sim\eta$, on the other hand,
 represents  the  environment-induced renormalization of the 
$x$ and $y$ components of the spin (g-factor). 
The above perturbative result and the survival of the Curie 
susceptibility indicates that the term $P_{-2G,G}(t)$ decays 
to a reduced but non-zero value for small $\eta$.

In contrast to the $x$ and $y$ components, the $z$ component of the
susceptibility, 
$\chi_z$, is found to remain unrenormalized by $\eta$ to leading order
in the dissipation.
These results imply that, for weak Ohmic dissipation, the
only effect of dissipation is to slightly and anisotropically 
renormalize the $g$-factor,
but apart from that the spin behaves as a free spin, and
does not decay.

\section{Case of $\beta\ne0$}

So far we discussed only the case
$\beta=0$.
We show now that the system with both $\alpha,\beta$ finite is
equivalent to the Hamiltonian (\ref{hpf}).
Assume a state $|q\rangle$ that is an eigenstate of ${\hat
  Q}|q\rangle=q|q\rangle$. This state generates a ladder of states,
$|m+q\rangle \equiv \eexp{i m \theta }|q\rangle $,
with integer $m$ by
\beq
{\hat Q}\eexp{im\theta}|q\rangle=(m+q)\eexp{im\theta}|q\rangle\;.
\eeq
Since ${\hat T}^{-1}{\hat Q}{\hat T}=-{\hat Q}$, a sequence of time
reversed states is also generated by the time reversal operator,
$\hat T$:
$
{\hat Q}{\hat T}|m+q\rangle=-(m+q){\hat T}|m+q\rangle\;.
$
%
All these states are orthogonal since they correspond to different
energy eigenvalues, and the environment couples the $m$ and $m\pm 1$ states, exactly as for
$\beta=0$. The only difference is that $E_{0\uparrow}(\alpha, \beta)$,
which is not known analytically, changes the factor
$\half-\half\sqrt{1+\alpha^2}$ in
$E_{m\uparrow},\,E_{m\downarrow}$. Hence Eq. (\ref{hpf}) is a correct
representation also of the $\beta\neq 0$ case.\footnote{The matrix
  elements of the spin operators are nevertheless different, 
changing e.g. the overall coefficient in Eq.~\eqref{pseudofermion_spin}.}

\section{Conclusions} 

We derived the effective Hamiltonian of an
electron confined to a ring within a 2-dimensional electron gas, in the
presence of SO coupling, and subject to a dissipative environment.
We have shown that the orbital motion of the particle decouples from
the spin evolution, and correspondingly, spin decay has a geometric
character~\cite{SanJose}. For an Ohmic environment,
we mapped the spin relaxation problem to that of a spinless particle on a ring
pierced by a magnetic flux [Eqs.~(\ref{C^x_++},\ref{C^z_++})]. 
We find that the $z$ component of the 
ground state spin is not affected by dissipation. 
The $x$ and $y$ in-plane spin components are, on the other hand,
 reduced by  dissipation,  but we find no dephasing for spin
$\half$ and weak dissipation. However, 
these components seem to dephase at large dissipation. 

We should remark that these latter results are based on the assumption 
of Ohmic dissipation. The situation may, however, change for 
subohmic dissipation or $1/\omega^\gamma$ noise, present in many 
systems.  In this case, 
the decoupling of the spin and orbital motion and thus 
Eqs.~(\ref{C^x_++},\ref{C^z_++}) remain valid, however, 
for subohmic  dissipation $\eta$ is a {\em relevant} perturbation, and 
even a small dissipation could possibly lead to the decay 
of the $x$ and $y$ spin components. This possibility, however, needs 
to be further explored.

\acknowledgments
We acknowledge useful discussions with B. Dou\c cot, W. Zwerger
and A. Zaikin. This research has been supported by
the Hungarian Research Funds OTKA and NKTH under Grant Nos.
K73361, 
T\' AMOP-4.2.1/B-09/1/KMR-
2010-0002,  and the EU-NKTH GEOMDISS project and by THE ISRAEL SCIENCE FOUNDATION (grant No. 1078/07).

\end{document}